\documentclass[apl,amsmath,amssymb,preprint]{revtex4-1}
\usepackage{graphicx}
\usepackage{dcolumn}
\usepackage{bm}

\begin{document}


\title[Sample title]{Selective growth of perovskite oxides on SrTiO$_{3}$ (001) by control of surface reconstructions}

\author{Soo-hyon Phark}
\email{phark@mpi-halle.mpg.de}
 \altaffiliation[present address: ]{Max-Planck-Institut f\"{u}r Mikrostrukturphysik, Weinberg 2, 06120 Halle, Germany}
\author{Young Jun Chang}
 \altaffiliation[present address: ]{Advanced Light Source, Lawrence Berkeley National Laboratory, Berkeley, California 94720, USA}
\author{Tae Won Noh}
\affiliation{ReCOE and FPRD, Department of Physics and Astronomy,
Seoul National University, Seoul 151-747, Korea}

\date{\today}

\begin{abstract}
We report surface reconstruction (RC)-dependent growths of
SrTiO$_{3}$ and SrVO$_{3}$ on a SrTiO$_{3}$ (001) surface with two
different coexisting surface RCs, namely (2$\times$1) and
c(6$\times$2). Up to the coverage of several layers, epitaxial
growth was forbidden on the c(6$\times$2) RC under the growth
conditions that permitted layer-by-layer epitaxial growth on the
(2$\times$1) RC. Scanning tunneling microscopy on the lattice
structure of the c(6$\times$2) RC revealed that this RC-selective
growth mainly originated from the significant
structural/stoichiometric dissimilarity between the c(6$\times$2)
RC and the cubic perovskite films. As a result, the formation of
SrTiO$_{3}$ islands was forbidden from the nucleation stage.
\end{abstract}

\pacs{Valid PACS appear here}
\keywords{Suggested keywords}

\maketitle

Thin film engineering via the atomic-scale control of surfaces and
interfaces is one of the most important technological assets for
regulating and utilizing the functionalities of oxide
materials.\cite{ohtomo2004, brinkman2007} In particular,
SrTiO$_{3}$ (STO) single crystalline substrates have been widely
used to realize low-dimensional oxide structures such as ultrathin
films\cite{chang2009} and nanowires,\cite{vasco2005} with
properties that are radically different from those of their bulk
hosts.

When utilized as a substrate for films composed of other
materials, a STO (001) surface terminated with a TiO$_{2}$ plane
and followed by the formation of a (2$\times$1) reconstruction
(RC) constitutes a well-defined surface structure.\cite{chang2010,
kawasaki1994, castell2002, erdman2002} However, slight changes in
temperature and/or oxygen partial pressure (\emph{P}$_{O2}$)
during surface preparation are frequently accompanied by a
substantial fraction of other type of RCs.\cite{castell2002,
jiang1999}

When growing films on STO (001) surfaces, growth mode and rate
often change in accordance with the terminations and/or RCs of the
substrate surfaces. These growth behaviors have been used to
obtain nanometer-sized patterned structures, such as arrangement
of molecules,\cite{deak2006} oxide nanowires,\cite{vasco2005} and
catalytic metal nanoparticles,\cite{silly2005} and control their
electrical properties. Such self-organized nanostructures show
great potential as an alternative, cost-effective, bottom-up
approach, but existing microscopic investigations of their
physical/chemical growth mechanisms are inadequate.

In this paper, we report scanning tunneling microscopy (STM)
observation on the growth behavior of STO and SrVO$_{3}$ (SVO)
films on STO (001) surfaces, on which (2$\times$1) and
c(6$\times$2)\cite{jiang1999, lanier2007} RCs coexist. In contrast
to (2$\times$1) RC, we found that the epitaxial growth of both
types of film was prohibited on c(6$\times$2) RC. Close STM
examination on atomic networks of c(6$\times$2) RC revealed that
the perovskite phase was not allowed from the nucleation stage.

Experiments were carried out with an STM system (base pressure $<$
2$\times$10$^{-10}$ Torr) combined with a pulsed-laser deposition
chamber (base pressure $<$ 2$\times$10$^{-9}$ Torr). We used
Nb(0.1$\%$)-doped STO (001) single crystals (CrysTec-GmbH) as
substrates. After a NH$_{4}$F buffered HF treatment, followed by
thermal annealing at \emph{P}$_{O2}$ = 1$\times$10$^{-2}$ Torr and
a substrate temperature (\emph{T}$_{sub}$) of 900$^{\circ}$C for
about 30 min, \emph{in-situ} STM and reflection of high energy
electron diffraction (RHEED) on the surfaces showed well organized
(2$\times$1) RC terraces separated by STO steps of one unit cell
(uc) (0.3905 nm) in height.\cite{kawasaki1994} To obtain the
c(6$\times$2) RC, we annealed the substrates at \emph{T}$_{sub}$ =
950$^{\circ}$C. In order to deposit STO and SVO films, a STO
single crystal and a sintered polycrystalline
Sr$_{2}$V$_{2}$O$_{7}$ target were ablated at \emph{P}$_{O2}$ =
1$\times$10$^{-4}$ Torr and \emph{T}$_{sub}$ =
600$-$700$^{\circ}$C using a KrF excimer laser ($\lambda$ = 248
nm) with a repetition rate of 1 Hz and an energy density of $\sim$
3 J/cm$^{2}$ on the target surfaces.

Figure 1 (a) shows the STO (001) substrate surface annealed at
\emph{P}$_{O2}$ = 1$\times$10$^{-2}$ Torr and \emph{T}$_{sub}$ =
950$^{\circ}$C for 60 min. The surface exhibits atomically flat
terraces. However, the profile along line A, shown in Fig. 1 (c),
reveals some regions half a uc ($\sim$ 0.2 nm) higher/lower than
normal terraces of the STO (001). The portion of such regions
increased with the longer annealing time.

Figure 1(b) is the atomic resolution STM image of such a region.
It clearly shows the stripe patterns running along [100] with the
protruding zigzag atomic networks. As shown in Figure 1(d), the
profile perpendicular to the stripe patterns (line B) revealed
large hill-valley structures. Note that the corrugation of $\sim$
0.1 nm is larger than the nominal atomic corrugations by the
factor of $\sim$ 10, and the stripe patterns clearly showed a
periodicity of 1.2 nm which is three times larger than the STO
lattice constant. We also confirmed this periodicity by two
additional spots\cite{koster2000} between the 0th and 1st
diffraction peaks in the RHEED pattern, as shown in the inset of
Fig. 1(b). Such structures have been identified as the
c(6$\times$2) RC of the STO (001),\cite{jiang1999, lanier2007} as
the uc is denoted by the rectangular box in Fig. 1(b).

Since STM has been known to image mostly Ti atoms for STO
surfaces,\cite{johnston2004} the characteristic features of the
c(6$\times$2) RCs observed in this study seem to originate from
the arrangement of Ti atoms. In addition, a model study on the STO
c(6$\times$2) RC predicted that the incorporations of Ti atoms
with different valences in the near-surface TiO$_{x}$ phases can
result in zig-zag ordering of truncated Ti octahedras at the
center of a stripe and considerable vertical corrugations of the
stripe patterns.\cite{lanier2007} Our observations are consistent
with those earlier works.

Figure 1(e) shows the power spectrum of the fast Fourier transform
(FFT) of Fig. 1 (b). The (1 0) and (1/6 1/2) spots represent the
(1$\times$1) and c(6$\times$2) surface structures, respectively.
Of greater interest is the existence of substructures, indicated
by the black mesh in Fig. 1 (b). This corresponds to the strong
(1/2 1/2) spots in Fig. 1 (d), representing c(2$\times$2)
structure.

To investigate the growth behavior of the perovskite oxides on
this surface of mixed RCs, we deposited a series of STO films, as
shown in Figs. 2(a) and (b). Nominal growths followed the
layer-by-layer mode. However, we observed disordered structures in
some regions, denoted by the black arrows in both images. A
magnified STM view (inset of Fig. 2(b))of such a region revealed
only the nano-sized grains created by random aggregation of
particle-like structures with sizes less than 1 nm. One may infer
that those regions were created from the deposition of STO on the
c(6$\times$2) RC by assuming the preservation of the c(6$\times$2)
RC portions on the substrate during growth.

Figure 2 (d) displays the profile along line A in Fig. 2 (b).
Aside from the irregular structures, some areas inside the
c(6$\times$2) RC region exhibit STM heights even lower than that
of the substrate surface, as indicated by the red arrows. Since
the STM height strongly depends on the electronic density of the
local position, the observation of such unphysical heights implies
that such regions should have different electronic properties from
those of the perovskite STO islands.

To understand the microscopic origin of those abnormal growth
behaviors, we examined how the island nucleates during the very
initial growth stage, realized by extremely small amount of STO
influx. We deposited STO by laser ablating the STO target with
only three pulses. Figures 2 (c) and (e) show the resulting
surface and the height profile, respectively, along the line B. As
indicated by the arrows, we observed nucleation of the STO
islands, with the well-known 1 uc height of $\sim$ 0.39 nm, in the
(2$\times$1) region. In the c(6$\times$2) region, however, we
could find no such island formation but defective structures of
the type shown in Fig. 2 (c). This observation clearly indicates
that the island growth was prohibited in the c(6$\times$2) RC
region of STO substrate from the nucleation process.

To investigate the role of surface RCs in heteroepitaxial growth,
we performed similar experiments on a SVO thin film. Figures 3(a)
and (b) show STM images of 0.7 and 1.2 ML SVO films, respectively.
The existence of islands with uniform 1-uc heights indicates the
layer-by-layer epitaxial growth of the SVO film. In contrast,
disordered structures in the regions around the steps can be seen
in both images. The inset of Fig. 3 (b) is a magnified view of
such a region. The underlying stripes along the [100] direction
verify that these regions correspond to c(6$\times$2) RCs. The SVO
films also formed disordered phases on the c(6$\times$2) RCs of
the STO (001) surface, as STO did on similar substrate surfaces.

When c(6$\times$2) RC and a (001) oriented perovskite film form an
interface, free-standing structures of both sides cannot be
preserved due to the large lattice mismatch arising from both
incommensurate substructure and large vertical deformation of the
c(6$\times$2) RC. Instead, growth could occur with unexpected film
orientations: for example, when CdTe film is grown on the STO, the
c(6$\times$2) RC permits the growth of (211) oriented films with
the low crystallinity, whereas the substrate of no RC allows the
(111) oriented films.\cite{neretina2009} However, this does not
fully explain the growth behaviors of STO and SVO films in our
cases, since no crystalline phase with preferred orientation
seemed to grow on the c(6$\times$2) RCs, as observed in the STM
images.

Near-surface stoichiometry of c(6$\times$2) RC could also play an
important role of prohibiting the formation of perovskite films by
inducing the electrical dissimilarity at the interface. Under
harsh preparatory conditions, such as prolonged sputtering and/or
high-temperature annealing, it is known that numerous titanium
oxide (TiO$_{x}$) phases and/or nanostructures can be formed on
STO (001) surfaces.\cite{deak2006, silly2004} Therefore, it is
quite plausible that the surface layer stoichiometry of the
c(6$\times$2) RC could be extremely different from that of the
ideal STO (001) termination.\cite{lanier2007} Additionally, it is
known that the c(6$\times$2) RC is quite chemically stable in both
O$_{2}$ and the air in spite of structural
complexities.\cite{jiang1999} Free-standing (001) STO or SVO
surfaces have surface electronic neutrality with uc periodicity.
However, the arrangement of the Ti atoms with different valences
(see Fig. 1 (b) and description in the text) in the c(6$\times$2)
RC would result in an electrical polarity modulation with a very
short periodicity at the interface. This charge modulation could
also produce a significant increase in the interface energy.

Therefore, the growth of the perovskite film on c(6$\times$2) RC,
even for that with a low textured orientation, would require large
excessive structural deformation and/or stoichiometry changes in
the interfacial region.

We demonstrated the significant influence of surface RCs on the
initial growth of perovskite epitaxial thin films by using the STO
(001) substrate, whose surface were composed of two RCs,
(2$\times$1), c(6$\times$2). The STM observations revealed that
the structural/stoichiometric dissimilarity between the
c(6$\times$2) RCs and the perovskite structure primarily
contribute to the selective growth of perovskite films only on
(2$\times$1)RC regions of such substrate. These observations
suggest the possibility of applying mixed RC surfaces to the field
of oxide heterostructure engineering.

The authors are grateful to M. R. Castell for valuable
discussions. This research was supported by the Basic Science
Research Program through the National Research Foundation of Korea
(NRF) funded by the Ministry of Education, Science and Technology
(No. 2009-0080567)

\bibliography{sto_c6x2_arxiv}

\begin{figure}
\includegraphics[width=10cm]{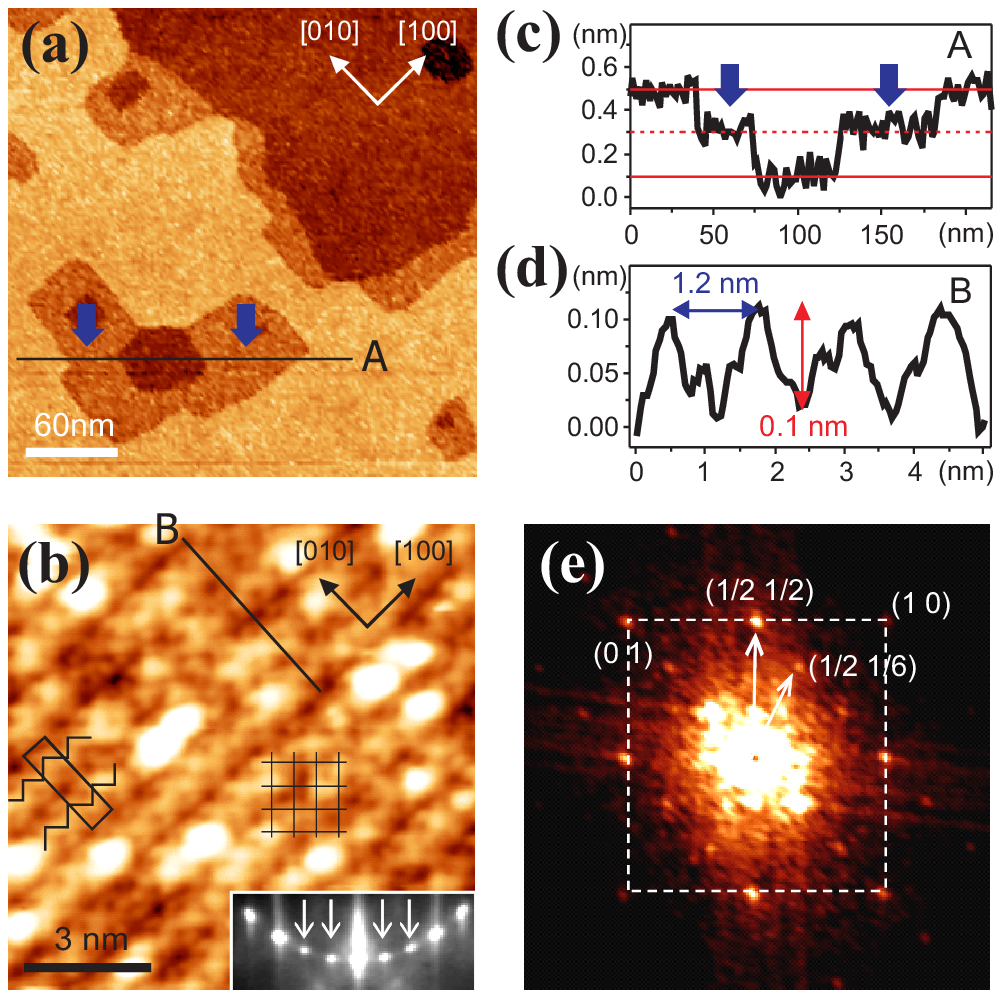}
\caption{\label{fig:epsart} (Color online) (a) and (c) show an STM
image of the substrate and the profile along line A, respectively.
The red solid (dotted) line denotes the vertical positions of
surface terminations with uc (half-uc) heights. (b) and (e) show
an STM image and its FFT power spectrum taken from the region
denoted by the blue arrows in (a). The inset of (b) displays RHEED
patterns taken from the substrate on which such regions covered
more than 50 $\%$ of the surface. (d) shows the profile along line
B. (\emph{V}$_{S}$ = 2.5 V and \emph{I}$_{set}$ = 50 pA)}
\end{figure}

\begin{figure}
\includegraphics[width=10cm]{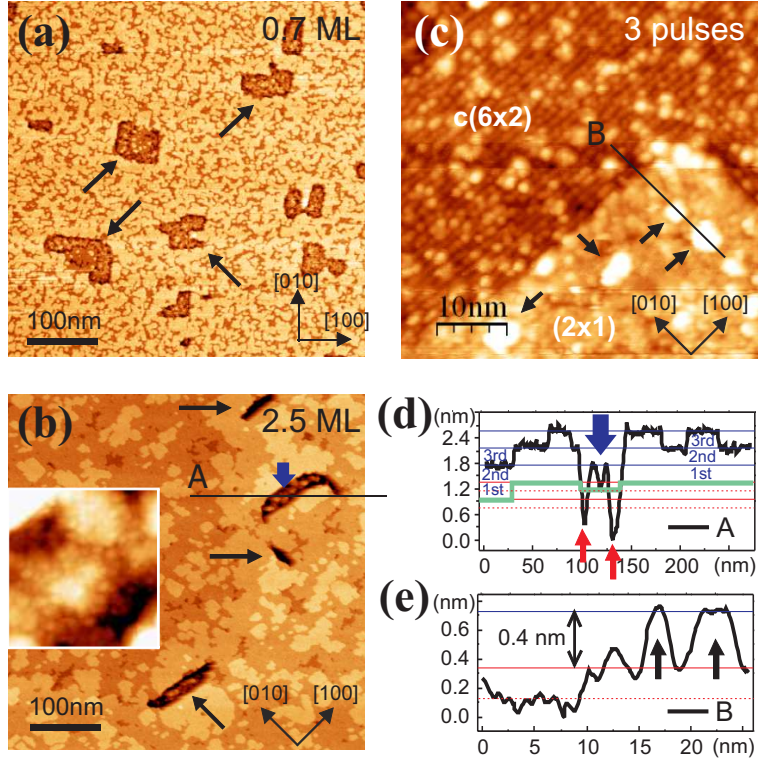}
\caption{\label{fig:epsart}(Color online) (a) and (b) show STM
images of 0.7 and 2.5 ML STO films. (c) is an STM image of the STO
(001) surface obtained by applying three laser pulses to the STO
target at \emph{P}$_{O2}$ = 1 $\times$ 10$^{-4}$ Torr and
\emph{T}$_{sub}$ = 600 $^{\circ}$C. (d) and (e) show the profiles
along lines A and B. Red solid(dotted) lines indicate the surface
terminations of the substrate with uc (half-uc) heights, and blue
lines show the heights of deposited STO layers. The sky-blue thick
line denotes the substrate surface. (\emph{V}$_{S}$ = 2.5 V and
\emph{I}$_{set}$ = 50 pA)}
\end{figure}

\begin{figure}
\includegraphics[width=10cm]{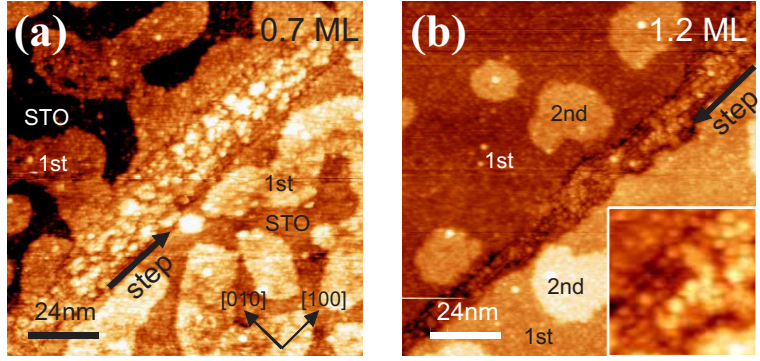}
\caption{\label{fig:epsart}(Color online) STM images of (a) 0.7 ML
and (b) 1.2 ML SVO films. `STO' denotes the surface of the
substrate. `1st' and `2nd' indicate the first and second layers of
the SVO film, respectively. The inset of (b) is a magnified view
of the region of disordered growth. (\emph{V}$_{S}$ = 2.5 V and
\emph{I}$_{set}$ = 50 pA)}
\end{figure}

\end{document}